\documentclass[12pt]{article}
\usepackage{graphics}
\def\cA{{\cal A}}
\def\cB{{\cal B}}

\def\cD{{\cal D}}

\def\cJ{\cal J}

\def\a{\alpha}

\def\b{\beta}

\def\d{\delta}

\def\m{\mu}
\def\n{\nu}

\def\r{\rho}
\def\s{\sigma}
\def\t{\tau}
\def\th{\theta}

\def\beq{\begin{equation}}\def\eeq{\end{equation}}
\def\beqa{\begin{eqnarray}}\def\eeqa{\end{eqnarray}}
\def\barr{\begin{array}}\def\earr{\end{array}}

\def\del{\partial}

\let\bm=\bibitem

\def\nn{\nonumber}
\def\bd{\begin{document}}
\def\ed{\end{document}}
\def\ba{\begin{array}}
\def\ea{\end{array}}
\def\bea{\begin{eqnarray}}
\def\eea{\end{eqnarray}}
\def\ft#1#2{{\textstyle{{\scriptstyle #1}\over {\scriptstyle #2}}}}
\def\fft#1#2{{#1 \over #2}}
\newcommand{\be}{\begin{equation}}
\newcommand{\ee}{\end{equation}}
\newcommand{\eq}[1]{(\ref{#1})}
\def\eqs#1#2{(\ref{#1}-\ref{#2})}
\def\det{{\rm det\,}}
\def\tr{{\rm tr}}
\newcommand{\ho}[1]{$\, ^{#1}$}
\newcommand{\hoch}[1]{$\, ^{#1}$}
\def\ra{\rightarrow}

\def\bk{{\bf k}}
\def\Xh{\hat{X}}
\def\ah{\hat{a}}
\def\xh{\hat{x}}
\def\yh{\hat{y}}
\def\ph{\hat{p}}
\def\G{{\cal G}}
\def\Dth{{\Delta_\th}}

\newcommand{\sect}[1]{\setcounter{equation}{0}\section{#1}}
\renewcommand{\theequation}{\thesection.\arabic{equation}}
\newcommand{\ex}[1]{{\rm e}^{#1}} \def\ii{{\rm i}}
\newcommand{\hepth}[1]{{\tt hep-th/#1}}
\newcommand{\hepph}[1]{{\tt hep-ph/#1}}
\newcommand{\req}[1]{Eq.~(\ref{#1})}
\def\mint{\int\limits}

\begin{document}
\thispagestyle{empty}

\hfill{NEIP-00-012}

\hfill{DFTT 17/00}

\hfill{hep-th/0004183}

\vspace{20pt}

\begin{center}

{\Large \bf Multiloop String Amplitudes with $B$-Field
and Noncommutative QFT}
\vspace{30pt}

{\large Chong-Sun Chu\hoch{1}, Rodolfo Russo\hoch{1},
Stefano Sciuto\hoch{2}}

\vspace{15pt}
\begin{itemize}
\item[$^1$]
{\small \em Institute of Physics, University of Neuch\^atel,
CH-2000 Neuch\^atel, Switzerland}
\item[$^2$]
{\small \em Dipartimento di Fisica Teorica, Universit\`a di
Torino;\\
I.N.F.N., Sezione di Torino, Via P. Giuria 1,
I-10125 Torino, Italy }

\end{itemize}

\vskip .2in \sffamily{chong-sun.chu@iph.unine.ch\\
rodolfo.russo@iph.unine.ch \\
sciuto@to.infn.it}

\vspace{50pt}
{\bf Abstract}

\end{center}
The multiloop amplitudes for the bosonic string in presence of a
constant $B$-field are built by using the basic commutation relations
for the open string zero modes and oscillators.
The open string Green function on the annulus is obtained
from the one loop scattering amplitude among $N$ tachyons. For higher
loops, it is necessary to use the so called three Reggeon vertex,
which describes the emission from the open string of another string
and not simply of a tachyon. We find that the modifications to the
three (and multi) Reggeon vertex due to the $B$-field only affect the
zero modes and can be written in a simple and elegant way. Therefore
we can easily sew these vertices together and write the general
expression for the multiloop $N$-Reggeon vertex, which contains
any loop string amplitude, in presence of the $B$-field. The field
theory limit is also considered in some examples at two loops and
reproduces exactly the results of a noncommutative scalar field theory.

\vspace{0.5cm}
April, 2000
\setcounter{page}0
\newpage


\sect{Introduction}

It was Einstein's remarkable idea that spacetime is dynamical
and can be deformed by energy momentum tensor, with  Riemannian
geometry being the appropriate mathematical framework.
It was realized recently that a constant $B$-field in string
theory can also deform
(D-brane worldvolume) spacetime \cite{DH,CDS}. This
kind of deformation is different from that of
general relativity.
In this case, the D-brane worldvolume
becomes noncommutative. See \cite{ncg1} for a
comprehensive introduction to noncommutative geometry.

The idea of noncommutative spacetime is not new.
However, unlike the early approaches, the current developments of
string theory display noncommutativity as a result of the underlying
string or M-theory and do not require any further input.
This has been clarified in
\cite{AAS,CH1,volker,sw} for the string theory case,
in \cite{CHL} for the M-theory case
and  has also been extended in \cite{dubna}
for the charged string case.

Different directions of investigation have been undertaken since
then. One direction is the study of the
perturbative aspects \cite{filk}-\cite{karl}
of noncommutative field
theory, with the important discovery of UV/IR mixing
\cite{seiberg1}. Another interesting direction   leaded by Alekseev,
Recknagel and Schomerus \cite{ARS1}
is the study of fuzzy physics for
D-branes in WZW models \cite{fuzzyd}.

There has also been a lot of interest recently
\cite{dorn,kl1,bcr,shenker,hong}
in exploring the relationships between
noncommutative field theory and string theory.
These studies have been partially motivated by the suggestion
of \cite{seiberg2} on the stringy origin of the UV/IR mixing,
However, 
in all these works the 1-loop
{\it open} string Green function is obtained from the
{\it closed} string Green function by restriction to the boundary.
This approach is indirect and there
is no convincing
reason that the closed string Green function should give rise
to the open string Green function in this manner.
One of the aims of the present paper is to settle this issue
using only open string theory without resorting to closed string or
comparison with field theory calculations.

In \cite{CH1} the quantization of an open string in the presence of a
$B$-field was carried out in the Hamiltonian formulation.
One advantage of this approach
is that  the commutation relations for the
modes of the string expansion are explicitly determined, with the
noncommutativity of the string coordinates obtained as a derived
concept. Therefore the results of \cite{CH1} are
particularly suitable for doing
calculations in the operator formalism, where
these commutation relations are essential.
As we will review below, turning on a $B$-field has the
effect of making the endpoints of open strings
noncommutative. In terms of the commutation relations of the
string expansion modes, it is  a simple  modification
to the zero mode commutation relation \eq{hcr3}, \eq{ycr3}.
This observation allows us to
construct the open string vertex operators with $B$-field and
determine the one-loop open string amplitudes directly. It is then
straightforward to extract the one-loop open string Green function.

For higher loops, the approach of
vertex operators becomes less useful. However the power
of the operator formalism remains. One can uniformly construct
higher loop string amplitudes by putting together (ie. sewing)
the basic building blocks. The basic building blocks in this formalism
are the BRST invariant open string propagator and three Reggeon vertex.
Intuitively it can be expected that turning on a $B$-field
does not modify the open string propagator. In fact, at field theory
level, it is easy to see that the quadratic part of the action is
unaffected by the noncommutative parameter and only the interaction
terms depend on it. We show that a similar pattern is present also at
string level. In this case, the interaction part can be summarized
by the three Reggeon vertex; the modification of this vertex
is again determined by the noncommutativity of the zero modes.
By sewing a suitable number of three Reggeon vertices by means of
propagators, we will build the generating functional for any $N$ point
amplitude, called  $N$-Reggeon vertex; we will find that, also in presence
of
$B$-field,   it can be
written down in an elegant and compact form at all loops.

The paper is organized as follows. In section 2, we construct the open
string vertex operators with $B$-field
and use them to compute the open string one loop amplitudes and
extract the one-loop open string Green function. In section 3, we
explain how $B$-field affects the basic building blocks in the
construction of multiloop string amplitudes and obtain the $h$-loop
$N$-point Reggeon vertex with $B$-field. In section 4, we consider the
noncommutative $\Phi^3$ field theory in six dimensions at the two loops
level and show that the field theory results
agree precisely with the field theory limit
of the two-loop string amplitudes we computed with
the Reggeon vertex.
We finally conclude with some open problems.


\sect{One-Loop Open String Green Function}

In this section, we introduce the vertex operators for open string
states in the presence of a constant $B$-field. Using these vertices,
we can compute string amplitudes among tachyons states at tree and
one-loop level; then, directly from this result, we extract the open
string Green function on the annulus. As was recently shown, this is
the basic ingredient for deriving noncommutative Feynman diagrams from
string theory~\cite{dorn}--\cite{hong}.

\subsection{Open String Vertex Operator}

We first recall the results \cite{CH1} for an open string
ending on a D-brane in presence of a constant $B$-field.
Here, we will need only the open string
coordinates in the longitudinal directions of the D-branes; in fact,
for the purpose of deriving field theory diagrams from string
amplitudes, it is sufficient to deal with spacetime filling D-branes. In
this background the open string mode expansion is
\be
X^\m(\tau,\sigma) =x_0^\m + 2\a' (p_0^\m \tau -  p_0^\n F_\n{}^\m \sigma) +
\sqrt{2\a'} \sum_{n\neq 0} {e^{-\ii n\tau} \over n}
\bigl(\ii a^\m_n \cos n\sigma -   a_n^\n F_\n{}^\m \sin n\sigma \bigr)~,
\label{mode1}
\ee
where $F =B-dA$ is the modified Born-Infeld field strength.
In order to write the commutation relations of these modes, it is
convenient to introduce the rescaled quantities
\be\label{resc}
\xh_0^\m = x_0^\n (1-F)_\n{}^\m, \quad
\ph_0^\m = p_0^\n (1-F)_\n{}^\m, \quad
\ah_n^\m = a_n^\n (1-F)_\n{}^\m~.
\ee
In fact, in terms of these operators, the equations take the standard
form, except for that of the zero modes \cite{CH1}
\bea
&[\ah_n^\m, \xh_0^\n]=[\ah_n^\m, \ph_0^\n] =
[\ph_0^\m,  \ph_0^\n] =0, \label{hcr1}\\
&[\ah_m^\m, \ah_n^\n]= m \eta^{\m\n}\d_{m+n,0}~,
\quad [\xh_0^\m, \ph_0^\n]=  \ii  \eta^{\m\n},\label{hcr2} \\
&[\xh_0^\m,\xh_0^\n]= 2\pi \ii \a'F^{\m\n} = \ii \th^{\m\n} ~.
\label{hcr3} \eea
In the last line we have introduced the dimensionful quantity $\th := 2
\pi \a' F$ which directly measures the noncommutativity. As we will
see, this simple modification of the commutation relations for the zero
mode $\xh_0$ plays a fundamental role in giving the whole
$\th$-dependence of the string amplitude. We also recall that the
normal-ordered Virasoro generators, defined in terms of the hatted
operators, satisfy the standard Virasoro algebra
(first paper in \cite{CH1})
with a central charge unmodified by $F$.

The vertex operators for open string states emitted at the boundary are
constructed using  $\Xh^\m$ evaluated at $\s=0$ or $ \s= \pi$,
\bea\label{e1}
\Xh^\m(\tau,0) &= & \xh_0^\m+ 2\a' \ph_0^\m \tau +
\ii \sqrt{2\a'}\sum_{n\neq 0} {e^{-\ii n\tau} \over n} \ah^\m_n~,
\\ \nonumber
\Xh^\m (\tau,\pi)  & = & \yh_0^\m +  2\a' \ph_0^\m \tau +
\ii \sqrt{2\a'}\sum_{n\neq 0} \frac{(-e^{-\ii \tau})^n}{n} \ah^\m_n~,
\label{Xhpi} \nn
\eea
where we have introduced
\be
\yh_0^\mu := \xh_0^\mu - 2\a' \pi \ph_0^\nu F_\nu{}^\mu~.
\ee
The zero-mode $\yh_0$, appearing in the expansion of  $\Xh^\m(\tau,\pi)$,
satisfies
\be \label{ycr2}
[\xh_0^\mu, \yh_0^\nu] = 0,
\ee
\be\label{ycr3}
[\yh_0^\mu, \yh_0^\nu] = -\ii \th^{\mu \nu} .
\ee
For instance, the open string tachyon vertex operator for emission at
$\s=0$ is
\bea \label{vtxopt}
V(p,\t) &=& : e^{\ii p \cdot \Xh(\t,0)}: ~  \\
&=& \exp\left(\sqrt{2\a'} \; p\cdot \sum_{n=1}^\infty
\frac{\ah_{-n}}{n}e^{\ii n \t} \right)
\; e^{\ii p\cdot\xh_0} e^{\ii 2\a' p \cdot\ph_0 \t}\;
\exp \left(-\sqrt{2\a'} \; p\cdot \sum_{n=1}^\infty
\frac{\ah_{n}}{n}e^{-\ii n \t}\right)~, \nn
\eea
while, for the emission at $\s=\pi$, we just have to replace $\xh_0$ by
$\yh_0$ in the above.
It is straightforward to check that it satisfies the desired property,
both for $\s=0$ and $\s=\pi$
\be
[L_m, V(p, \t) ] = e^{i m \t} (-\ii \frac{d}{d \t} + \frac{1}{2} m p^2)
V(p, \t) ~.
\ee
Similarly, one can construct the vertex operators for higher open
string states. For example, the gluon vertex operator is
\be \label{vtxopg}
V_{\varepsilon}(p,\tau) =  \varepsilon \cdot \frac{d \Xh}{d \t} ~
e^{i p\cdot \Xh}  \; .
\ee

We want to stress that the commutation relation \eq{ycr3} has an
opposite sign compared to \eq{hcr3}. This is in agreement with the
result of \cite{CH1} where it was shown that
\be \label{Xpm}
[\Xh^\m(\t,\s),\Xh^\n(\t,\s')] = \left\{
\begin{array}{ll}
\pm i \th ^{\m\n}, & \s=\s' =0 \mbox{ or } \pi,\cr
0, &  \mbox{otherwise},
\end{array}
\right.
\ee
This means that the noncommutativity of the string
coordinates $\Xh^\m$ at equal time is entirely determined by the zero
modes.
Note that \eq{Xpm} is equivalent to say that multiplication for
functions defined on the D-brane worldvolume are done
in opposite ordering
(sec. 5 of first paper in \cite{CH1} and sec. 6.2 of \cite{sw}).
This difference in the commutation relation for $\xh_0$ and $\yh_0$
is not so important for tree level calculations, since one can choose to
put all interactions at $\s=0$. However, this difference becomes
essential for one-loop diagrams: in fact, in this case non-planar
diagrams require to put vertex operators at the two different
borders, $\s=0$ and $\s=\pi$.

\begin{figure}[ht]
\begin{center}
{\scalebox{1}{\includegraphics{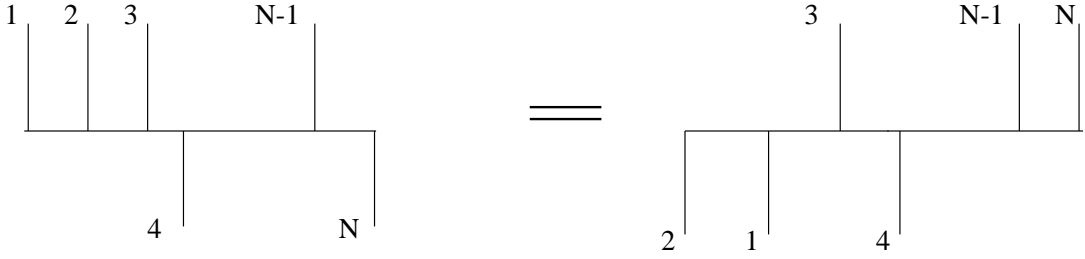}}}
\end{center} \caption{Cyclicity of the tree level $N$-Reggeon vertex.}
\label{cyclic}
\end{figure}

Finally, we remark that the vertex operator $V_\a(p,\tau)$ for a string
state $\a$ satisfies
\be\label{perm1}
V_{\a_1}(p_1,\t_1) V_{\a_2}(p_2,\t_2) =
V_{\a_2}(p_2,\t_2)V_{\a_1}(p_1,\t_1) e^{i \phi} ,
\ee
where
\be \label{perm}
\phi = \left\{
\begin{array}{c@{}cl}
& 2\pi \a'\; p_1\cdot p_2\;  \epsilon(\t_1-\t_2) - p_1\,\th\, p_2~,
& \mbox{if $\a_1$ and $\a_2$ are  on the same border $\s=0$},
\cr
& - 2\pi \a'\; p_1\cdot p_2\;  \epsilon(\t_1-\t_2)+p_1\,\th\, p_2~,
&
\mbox{if $\a_1$ and $\a_2$ are  on the same border $\s=\pi$},\cr
&0,     &\mbox{if $\a_1$ and $\a_2$ are  on different borders, }
\end{array}
\right.
\ee
and $\epsilon (\t) = \t/|\t|$.
The $\th$-dependent piece  is a simple consequence of the zero mode
commutation relations \eq{hcr3}, \eq{ycr2} and \eq{ycr3} and of the fact
that
either the zero
mode $\xh_0$ or $\yh_0$ appears in the vertex, depending on whether it
is on the $\s=0$ or $\s=\pi$ border. The
difference of sign for the two borders in \eq{perm} is crucial, since it
ensures that tree-level string amplitudes are invariant under cyclic
permutations. For instance, it is easy to work out the $N$-point
amplitude, where some of the legs are emitted from the border $\s=0$ and
the others from $\s=\pi$, and then to verify that the result is
cyclically invariant: one has simply to use the
commutation relation \eq{perm1} and momentum conservation.
In Fig \ref{cyclic}, vertices
inserted on the $\s=0$ ($\s=\pi$)  boundary are represented by a
vertical line above (below) the horizontal line.

\subsection{Open String Green Function}

\begin{figure}[ht]
\begin{center}
{\scalebox{1}{\includegraphics{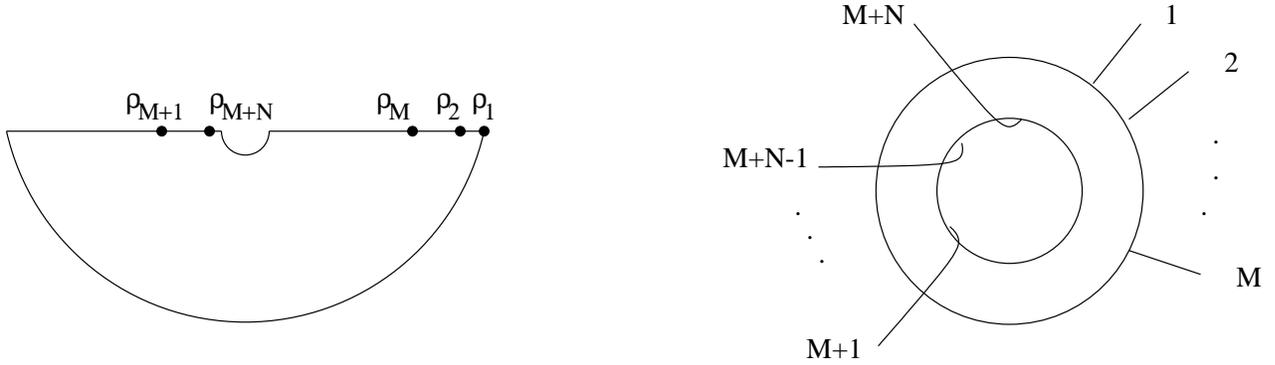}}}
\end{center} \caption{One annulus in Schottky representation.}
\label{scho1loop}
\end{figure}

Now we compute the 1-loop open string amplitude for tachyons
using the vertex operators constructed above and extract the open
string Green function from it.
Consider the $M+N$-point amplitude depicted in Fig.~\ref{scho1loop}.

The external states $n \in \{1,
\cdots, M \} :=I_1$ are on the boundary $\s=0$, while the other
particles $ s \in \{M+1, \cdots, M+N \} :=I_2$ are
on the boundary $\s= \pi$; moreover, we have chosen the radial
ordering
\be \label{config3}
k \leq |\r_{M+N}| \cdots \leq |\r_{M+1}|
\leq \r_M \cdots \leq \r_2 \leq \r_1 \leq 1.
\ee
The 1-loop amplitude is given by
$$
A(1,\ldots, M; M+1, \ldots,M+N) \sim 
\int d^d p \; \tr \Big\{
V(p_1,\rho_1) \ldots V(p_{M+N},\rho_{M+N})\, k^{L_0}
\Big\},
$$
where
$V$ is the vertex introduced in \eq{vtxopt}.
First we look at the part containing $\xh_0$ and $\yh_0$, which is given
by
\be \label{zeromode}
e^{\ii p^1 \xh_0} \ldots\, e^{\ii p^M \xh_0}
e^{\ii p^{(M+1)} \yh_0 }
\ldots\, e^{i p^{(M+N)} \yh_0  } \nn\\
= e^{ -\ii \ph_0 \th \sum_{s \in I_2} p^{s}}
e^{-\frac{i}{2} \sum_{\stackrel{r <s}{r,s \in I_1}  } p^{r}\th p^{s}}
e^{ \frac{i}{2} \sum_{\stackrel{r <s}{r,s \in I_2 }} p^{r}\th p^{s}}~.
\ee
Evaluating the trace turns $\ph_0$ into the loop momentum $p$.
Notice that, besides the usual $p$ dependent factors coming from the
propagators, one has a new term linear both in $\th$ and in $p$
which has its origin in the commutation relations of the zero
modes $\xh_0$ and $\yh_0$~\footnote{The
importance of the commutation relations for the zero modes was
emphasized, in a tree-level example, in
the first paper of \cite{susy}. There one has $\ph_0=0$ and thus
obtains the usual Filk phase.}. Globally the zero mode part becomes
\bea \label{gaussian}
\int d^d p \, \exp
\left(
\a' p^2 \ln k  + p \cdot
\Bigg(2\a' \!\!\sum_{r\in I_1\cup I_2}\!\! p^{r} \ln |\r_r|
-\ii \th \sum_{s\in I_2} p^{s}\Bigg)
- \a'\!\!\!\!\!   
\sum_{\stackrel{r<s}{r,s \in I_1\cup I_2}}\!\! \!
p^{r} \cdot  p^{s} \ln \left|\frac{\r_s}{\r_r}\right|
\right) .
\eea
The Gaussian integration over $p$ is performed, as usual, by completing
the square; in particular, the expression in the
exponent is equal to
$$ \a' \ln k \left[ p + \frac{1}{\ln k}
\Bigg(\sum_{r\in I_1\cup I_2} p^{r} \ln |\r_r|
-\ii\frac{\th}{2\a'}\sum_{s\in I_2}p^{s}\Bigg)\right]^2
+  \a' \!\!\!\!\!
\sum_{\stackrel{r<s}{r,s \in I_1\cup I_2}}\!\!\!
p^{r} \cdot  p^{s}
\Bigg( \frac{ \ln^2 |\r_s /\r_r|}{ \ln k}
- \ln \left|\frac{\r_s}{\r_r}\right|\Bigg)
$$
\bea \label{xxx}
+ \frac{1}{\a' \ln k}\sum_{r\in I_1,s\in I_2}
p^{r}\left(\frac{\th}{2}\right)^2
p^{s} + \frac{\ii}{\ln k} \sum_{r\in I_1\cup I_2}\; \sum_{s\in I_2 }
\ln |\r_r|  p^{r} \th  p^{s}~.
\eea
It is easy to see that the last term in the above equation is equal to
\be \label{xxx1}
\frac{\ii}{ \ln k}
\sum_{r\in I_1, s\in I_2} \ln |{\r_r}{\r_s}|  p^{r} \th  p^{s}.
\ee
The momentum integral can now be evaluated and the
second line of \eq{xxx} contains two new terms arising in
nonplanar diagrams. In the non zero-modes part, the extra minus sign
in front of the $\ah_n$ oscillators in the $\Xh(\pi)$ expansion
has the effects to change the sign of $\r$ in the oscillator sum
for the nonplanar vertices.
It is worthwhile to remark again  that the $\th$ and $\th^2$
terms above arise as a result of the new linear $p$-shift in the
Gaussian integration.
We will see that the same pattern repeats
for the general case of the $h$-loop Reggeon vertex.

Finally we can summarize 1-loop amplitudes in the following compact
equation
\bea \label{A-mn}
A (1, \!\!\!\! &\ldots,&\!\!\!\! M; M+1, \ldots, M+N) \sim 
\int\limits_0^1 \frac{dk}{k^2}\left(\frac{-2\pi}{\ln k}\right)^{d/2}
(1-k^n)^{2-d}
\int \frac{d\r_2}{\r_2} 
\cdots \frac{d\r_{M+N}}{\r_{M+N}} \nn \\
&&\prod_{\stackrel{r <s}{r,s \in I_1}}
 e^{2 \a' G_P^{\m\n}(\r_r,\r_s)  p^{r}_\m  p^{s}_\n }
\prod_{r \in I_1, s\in I_2}
e^{2 \a' G_{NP}^{\m\n}(\r_r,\r_s)  p^{r}_\m  p^{s}_\n }
\prod_{\stackrel{r <s}{r,s \in I_2}}
e^{2 \a' G_P^{\m\n}(\r_r,\r_s)  p^{r}_\m  p^{s}_\n }~,
\eea
where the
planar and nonplanar open string Green function are
\be \label{p1}
G_P^{\m\n}(\r,\r') = I_0^P \eta^{\m\n} -  \frac{\ii\th^{\m\n}}{4 \a'}
\epsilon(\r-\r')\; ,\!
\quad
G_{NP}^{\m\n}(\r,\r') = I_0^{NP} \eta^{\m\n}
+ \frac{(\th^{2})^{\m\n}}{ 8 \a'^2}\frac{1}{\ln k}
\pm  \frac{\ii\th^{\m\n}}{2 \a'} \frac{\ln|\r \r'|}{\ln k},
\ee
with the $\th$-independent piece $I_0$ given by
\beq \label{p2}
I_0^P(\r,\r') = \frac{ \ln^2 \r / \r'}{2 \ln k} +
\ln \left|\sqrt{\frac{\r}{\r'}} - \sqrt{\frac{\r'}{\r}}\right|
+  \ln \prod_{n=1}^{\infty}
\left|\frac{(1-k^n \r/\r')(1-k^n \r'/\r) }{(1-k^n)^2} \right|  
\eeq
for the planar case, while for nonplanar contractions one has
\beq
I_0^{NP}(\r,\r') = \frac{ \ln^2 |\r / \r'|}{2 \ln k} +
\ln (\sqrt{\frac{|\r|}{|\r'|}} + \sqrt{\frac{|\r'|}{|\r|}} ) + \ln
\prod_{n=1}^{\infty} \left|\frac{(1+k^n |\r/\r'|)(1+k^n |\r'/\r|)
}{(1-k^n)^2} \right|~.
\eeq
In the last term of the nonplanar Green function in \eq{p1},
positive sign is taken when $\r >0, \r' <0$ and negative sign is taken
for the opposite case  $\r' >0, \r <0$. This follows directly from
our result \eq{xxx1}.

By using these Green functions, one can compute different string
amplitudes that, in the limit $\a'\to 0$, reproduce the Feynman diagrams
of diverse noncommutative field theory.
For example, by introducing in \eq{A-mn}
the Schwinger parameters $t_i~ (2\leq i \leq M+N$)
\be
- \a' \ln |\r_k| = t_2+ t_3+ \cdots +t_k, \quad 2\leq k \leq M+N,
\ee
and  taking $\a' \rightarrow 0$
with  $\th = 2\pi \a' F$ and $t_i$ both kept fixed,
the result is precisely the same as
the corresponding noncommutative $\Phi^3$
field theory amplitude, derived by means of the
noncommutative Feynman rules. This provides an independent
{\it field} theory check of the correctness of the string computation
\eq{A-mn} and of the  vertex operators we constructed above.
By contracting some of the propagators \cite{bcr},
one can also obtain  from \eq{A-mn} other field theory limits,
for example noncommutative $\Phi^4$.

We note that, in the
tachyon amplitudes \eq{A-mn}, the Green functions
are always contracted with the external momenta. Because of
this, it is also possible, by using momentum conservation, to rewrite
the last term of \eq{xxx} as a
function of the ratio of the $\r$'s over the two borders,
\be \label{lastterm}
- \frac{\ii}{\ln k} \left(
\sum_{\stackrel{r<r'}{r,r'\in I_1}}
\ln \left|\frac{\r_r}{\r_{r'}}\right| p^{r} \th  p^{r'}
-\sum_{\stackrel{s<s'}{s,s'\in I_2}}
\ln\left|\frac{\r_s}{\r_{s'}}\right| p^{s}\th  p^{s'} \right)~.
\ee
This shows explicitly that the tachyon amplitude is invariant under the
transformation
\be
\r_s \to \lambda \r_s, \quad \mbox{and} \quad \r_s \to -\lambda/\r_s,
\quad \lambda >0 .
\ee
It is easy to generalize the manipulations here and to show that this is
also true for the gluon amplitudes as well as the amplitudes for other
higher string states.

We also remark that
written in the form \eq{lastterm},
one may shift this piece to the planar diagram
and extract a different planar and nonplanar Green function
\be \label{wrong}
G'^{\m\n}_P = I_0^P \eta^{\m\n} -  \frac{\ii\th^{\m\n}}{4 \a'}
\epsilon(\r-\r') \mp
\frac{\ii \th^{\m\n}}{2 \a' \ln k}
\ln \left|\frac{\r}{\r'}\right|, \quad\quad
G'^{\m\n}_{NP} = I_0^{NP} \eta^{\m\n}
+ \frac{(\th^{2})^{\m\n}}{ 8 \a'^2}\frac{1}{\ln k} .
\ee
where $-$ sign is for the positive border and $+$ sign is for the
negative border.
Incidentally, these are the same as one obtained from
{\it restricting} to the boundary of
the 1-loop {\it closed} string Green function \cite{dorn}.
However, as stressed in \cite{bcr}, \eq{p1} and \eq{wrong} are not
equivalent when one wants to deal with external massless states and
derive, for instance, the Feynman diagrams of the noncommutative
gauge theory.
Indeed the necessity
of shifting the result naturally obtained from the closed
string calculations is clear within string theory, without
the need to refer
to any field theory results.
For instance, by following the same
procedure of this section, one can compute a general 1-loop gluon
interaction if the vertex operator \eq{vtxopg} are used instead of
\eq{vtxopt}. Then one has to rewrite the final amplitude in the form of
the string master formula for gluons, and identify the Green function
from it. The result is indeed given by \eq{p1}, \eq{p2} above,
naturally derived also in the scalar computation.

\sect{Reggeon Vertex and Multiloop String Amplitudes}

In this section we first construct the tree-level BRST invariant
$N$-Reggeon vertex for open
bosonic string in presence of a constant $B$-field. This object is the
generator of all tree-level scattering amplitudes among arbitrary string
states and will provide the first basic ingredient for constructing
higher loop ``noncommutative'' string diagrams. It is quite clear why
it is necessary, for our purposes, to introduce the Reggeon vertex
formalism. As usual, in order to construct loop amplitudes, one sews
together pairs of external legs and sums over all contributions coming
from the exchange of the different string states in the various loops. In
the previous section this has been achieved by simply identifying the
first ($z=\infty$) and the last ($z=0$) leg and then taking the trace.
This approach is no longer adequate for constructing higher loop
amplitudes; in fact, in order to repeat the sewing procedure, we need to
identify further couples of legs that cannot be fixed at
$z=0$ and $z=\infty$. However, in the vertex operator formalism,
all these legs have been represented by vertex
operators describing specific string states and thus it is very
difficult to {\it sum}
the contributions coming from the exchange of different
states. As we will briefly review,
the introduction of Reggeon vertex solves
this problem.

The second fundamental ingredient that is necessary for constructing
higher loop interactions is the BRST invariant open string propagator.
From the geometrical point of view, this propagator is a conformal
transformation identifying two circles around the punctures sewn
together. It turns out that this ingredient is not modified by the
presence of a constant $B$-field. Thus the $B$-dependent $N$-Reggeon
vertex will be the only new building block we need to use for computing
the ``noncommutative'' string amplitudes at loop level.


\subsection{Tree level Reggeon Vertex}

The starting point is a generalization of the usual vertices describing
the emission of string states. As is clear by looking at the simplest
case
\eq{vtxopt}, these vertices are built by means of the coordinates of the
propagating virtual string and their explicit form is determined by
the quantum numbers of the emitted states like momentum
or polarizations. In the generalization of these usual vertices, one
basically
replaces~\cite{3reg}
the quantum numbers of the emitted string with
a whole Hilbert space.
Thus, considering also the ghost contribution, the $3$-Reggeon vertex
is~\cite{sc0}~\footnote{The ghost contribution to the symmetric vertex
is given in~\cite{nw}}
\be \label{V30}
V_{3;0}^\th(\zeta) = \langle 0; q=3|
: \exp\left\{\oint_0 dz (-X^v(\zeta+z)\partial_z X(z)-
c^v(\zeta+z)b(z) +b^v(\zeta+z) c(z) )\right\}:~,
\ee
where the bra indicates the vacuum of the emitted string
for both the oscillators and the zero
mode $x_0$, while the label $q=3$ specifies the ghost number
\be
|0;q=3\rangle = | x_0=0;0_a\rangle \otimes c_{-1} c_0 c_1 | q=0\rangle ~.
\ee
Besides the fields of the virtual string denoted by $v$, in the
$3$-Reggeon vertex, also the coordinates of the external string appear
directly; the commutations relations of these new modes are the usual
ones \eq{hcr1}--\eq{hcr3}, while the oscillators of the virtual and the
external strings simply commute among them. Formally our expression
is identical to the one normally used in string calculation in the
trivial background $B=0$; however, \eq{V30} depends on the value of the
$B$-field through the mode expansion \eq{e1} and the new commutation
relations, which have to be used both for the virtual and the external
string. Here for convenience, we have dropped the hat over the $X$
fields, although it should be understood that all the $X$'s above contain
the oscillators defined in \eq{resc}. Notice that, in terms of the
rescaled coordinates, the zero-mode $x_0$ (or $y_0$ if the interaction
is at $\s=\pi$) appears only in the expansion of the virtual string and
is the only source of the non-trivial dependence on $\theta$.
Together with the Reggeon vertex \eq{V30}, we will employ the usual
physical states having ghost number $1$; thus, the $(b,c)$ system is not
affected by the background field $F$ and one recovers the well known
results for ghosts. Because of this, in what follows, we will no longer
mention ghosts and focus only on the $F$-dependent modifications coming
from the orbital part.

The key property of the 3-Reggeon vertex is that it can be saturated
with a string state $\a$ and one recovers the usual vertex operator
corresponding to the considered state
\be
V_{3;0}^\th (\zeta)|p;\a \rangle = V_\a (p,\zeta)~.
\ee
It is easy to check that when completely saturated with external states,
the 3-Reggeon vertex gives correctly the tree level amplitude, including
the phase factor
\be
{_v}\langle p_1;\a_1|~ V_{3;0}^\th~ |p_2;\a_2\rangle \;|p_3;\a_3\rangle_v
\sim {\rm e}^{\mp\frac{i}{2} p^1\th p^2}~,
\ee
where, again, the upper (lower) sign refers to the emission of the
particle $\a_2$ from the border $\s=0$ ($\s=\pi$) of the propagating
string $v$; in particular, the phase arises because of the use of the
commutation relation \eq{hcr3} (or \eq{ycr3}) for the zero modes.

We remark that all the formula written in this section are
general and hold for any value of $\a'$. In the next section, we will
consider the noncommutative field theory limit
\be \label{double}
\a' \to 0, \quad F \to \infty
\ee
with $\a' F$ fixed so
that a noncommutative field theory is obtained.

The tree level $N$-Reggeon vertex can be constructed simply by
multiplying $N$ $3$-Reggeon vertices in different positions $\zeta$, but
written in terms of the same propagating string $v$. Then, one can
explicitly evaluate the vacuum expectation value in the Hilbert space
$v$ in order to keep only the dependence on the oscillators of the
external states. The new $\th$-dependent part comes when one collects
together the zero mode factors. In particular, if the external legs of
all the original 3-Reggeon vertices are emitted from the border $\s=0$,
one obtains the new phase factor
\be
e^{i p^1 x_0} \cdots e^{i p^N x_0} =
e^{-\frac{i}{2} \sum_{r <s}^N {p}^{r}\th {p}^{s}},
\ee
where momentum conservation has been used.
As a result, we obtain the $N$-Reggeon vertex with all legs
emitted from the $\s=0$ border
\be \label{tree}
V_{N;0}^\th =  V_{N;0}^0 \exp (-\frac{i}{2}
\sum_{i<j}^N {p}^{i} \th {p}^{j}),
\ee
where ${p}^{i}$ is the momentum {\it  operator} of the $i$-th leg, in the
direction flowing towards the boundary. Here $V_{N;0}^0$ indicates the
$N$-Reggeon vertex derived for the usual ``commutative'' case $F=0$ in
\cite{sc2,sc3}.
This vertex is a bra in the direct product of the $N$
distinct Fock spaces for the external strings; like the $3$-Reggeon
vertex, it gives the scattering amplitude when the external legs are
saturated with physical states.
In particular, like the 3-Reggeon vertex,
$V_{N;0}^0$ contains the vacuum of the zero modes
\be \label{vac}
\prod_{i=1}^N \; \langle x_0^i =0 |
= \prod_{i=1}^N \int d p^i \; \langle p^i | .
\ee
The momentum operator dependent phase factor in \eq{tree}
is to be interpreted as an
operator acting on the R.H.S. of \eq{vac}. When saturated with
external states, it gives a (numerical)
phase factor as the $p^i$'s become
the external momenta. In the next section, we will sew together legs
of the tree level Reggeon vertex \eq{tree}
to construct the multiloop Reggeon
vertex. We will see that the momenta of the corresponding legs are to
be identified and interpreted as the loop momentum.
\eq{vac} effectively instructs us to integrate over the loop momentum.

We also note that the only modification to the tree level $N$-point
vertex is the momentum dependent phase factor. This
is exactly the same kind of modification to the tree level vertex in a
noncommutative field theory due to the Moyal $*$-product.
We will see that this identification of the modifications in the string
vertex and field theory vertex is the basic reason which guarantees
that the usual one-one correspondence between corners of string moduli
space and field theory Feynman diagrams continues to hold in the
noncommutative case.

\subsection{ $h$-loop $N$-Reggeon Vertex}

The $h$-loop $N$-Reggeon vertex in the presence of a constant $F$-field
can be constructed by sewing together legs of a tree level
($N+2h$)-Reggeon vertex.
For example, in order to sew together leg $2$ and leg $(N-1)$,
we identify the corresponding Fock spaces
\bea
& a_n^{(2)} \to a_n^{(N-1) \dagger}, \quad  p^{(2)} \to - p^{(N-1)}, \\
& b_n^{(2)} \to  b_n^{(N-1) \dagger},
\quad\quad c_n^{(2)} \to  -c_n^{(N-1) \dagger}.
\eea
The next step in the sewing procedure is to identify the punctures
of the legs
that are going to be sewn together. In terms of local
coordinates $V_i^{-1}$ around the puncture $z_i$,
this means we have to identify $V_i^{-1}(z)$ and $V_j^{-1}(z)$
so that a complex structure can still be defined. In \eq{V30}, we have
chosen for the sake of simplicity
\footnote{
In taking the field limit, a more convenient choice
$ V_i(z)= z_i z+z_i$
is usually taken \cite{russo1}.
This has the effect of shifting the Green function
by the term  $-\frac{1}{2} \ln |z_i z_j|$
so that the resulting Green function $G(z_i,z_j)$
is of conformal weight $(-1/2,-1/2)$ in $(z,z')$ .
Off shell continuation of string results is more convenient using
this Green function. This choice is also adopted in the recent
analysis \cite{bcr}.
}
$V_i(z)= z+z_i$.
However, the
$N$-Reggeon vertex can be written for general
conformal transformation  $V_i$'s and it is easy to check that
the string amplitude is independent of the choice of them.
Because of this arbitrariness in the choice of the $V_i$,
there are in general many more redundant variables in the
Reggeon vertex than in the physical amplitudes. It is therefore
useful to make some convenient choices on the $V_i$'s so as to obtain
a simpler expression for the Reggeon vertex in order to
simplify intermediate calculations and manipulations.
In particular,
for constructing higher loop amplitudes through the sewing procedure,
the Lovelace choice \cite{lovelace} is  useful. In this choice, $V_i$
is a projective transformation which maps $\infty,0,1$ to $z_{i-1},
z_{i}, z_{i+1}$.
The sewing is achieved by using the BRST invariant sewing operator
$P(x)$ \cite{sc3}, which is a function of $L_0$ and $L_{\pm 1}$ and of the
ghosts.
Since $L_n$ does not involve the zero modes $x_0$, we conclude that the
string propagator $P(x)$ and the sewing procedure are not modified by
the presence of $F$-field. Therefore the only new feature for
$\th  \neq 0$ is in the zero modes part of the Reggeon vertex.

\begin{figure}[ht]
\begin{center}
{\scalebox{1}{\includegraphics{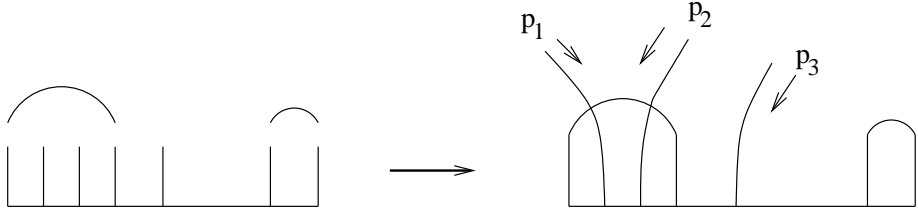}}}
\end{center} \caption{A Reggeon vertex with external
momenta $p_1, p_2$  crossing an internal momentum.}
\label{sewloop}
\end{figure}

Fig \ref{sewloop} shows how to construct a $N$-Reggeon vertex with all
legs emitted on the border $\s=0$.
As we have already said the ghost system is not modified by the
$F$-field,  thus its contribution (both zero and nonzero modes) to the
Reggeon vertex is unchanged.
Using \eq{tree} and recalling \eq{vac}, it is straightforward to
show that the $h$-loop $N$-Reggeon vertex in the presence of $F$-field
is
\be\label{Vloop}
V_{N;h}^\th = \widetilde{V}_{N;h}^{0}
\prod_{I=1}^{h} \int d p^I \exp
\left(
\frac{1}{2} \sum_{I,J =1}^h p^I_\m A_{IJ}^{\m\n} p^J_\n +
\sum_{I =1}^h B_I^\m p^I_\m + C
\right),
\ee
where $\widetilde{V}_{N;h}^{0}$ contains the ghost contribution and
only the nonzero mode piece of the orbital part of $N$-Reggeon vertex
in absence of background. Its explicit form can be found in
\cite{sc1,sc2}. What interests us is that all the dependence in $\th$
is
localized in the zero modes loop momentum integration.
We obtain the following  modifications to $A,B,C$
\bea
&A_{IJ}^{\m\n} ={A^0}_{IJ}^{\m\n} - i \th^{\m\n}\cJ_{IJ}  \label{a},\\
&B_I^\m = {B^0}_I^\m - i \th^{\m \n} {P}_{I\n},\label{b} \\
&C= C^0 - \frac{i}{2} \sum_{i<j}^N {p}^i \th {p}^j,  \label{c}
\eea
where $p_I$, $I=1, \cdots h$ are  the loop momenta; $p_i$, $i= 1,\cdots,
N$ are the external momenta;
$\cJ_{IJ}$ is the intersection matrix for the internal loops;
$P_I$ is the sum of external momentum entering the $I^{th}$ loop.
For example in Fig \ref{sewloop}, it is
\be
P_1 = p_1+p_2, \quad P_2 =0.
\ee
The explicit form of $A,B,C$ for $\th=0$ are given in \cite{sc1},
\bea
{A^0}_{IJ}^{\m\n} &=&  2\a'(2\pi i \t_{IJ}) \d^{\m\n} ,\\
{B^0}_I^\m &=&
\frac{1}{2\pi}  \sum_{i=1}^N \oint_{0} dz \del X^{(i) \m}(z)
\int_{z_0}^{V_i(z)} \omega_I ~ ,\\
C^0 &=& -\frac{1}{2}\sum_{i=1}^N
\oint_{0} dz \del X^{(i)}(z) p_0^{(i)} \ln V'_i(z) \\
&&+ \frac{1}{2\a'} \sum_{i<j}^N \oint_{0} dz \oint_{0} dy \del X^{(i)}(z)
\ln [V_i(z)-V_j(y)] \del  X^{(j)}(y) \nn\\
&&+ \frac{1}{4\a'} \sum_{i,j=1}^N \oint_{0} dz \oint_{0} dy
\del X^{(i)}(z) \ln \left(\frac{E(V_i(z),V_j(y)) }{V_i(z)-V_j(y) }\right)
\del  X^{(j)}(y) . \nn
\eea
Here $V_i(z)$ is chosen to satisfy $V_i^{-1}(z)=0$ for $z=z_i$.
$\omega_I$ is the normalized Abelian
differential,
\be
\oint_{a_J} \omega_I = \d_{IJ}
\ee
and
\be
2 \pi i \t_{IJ} = \oint_{b_J} \omega_I
\ee
is the period matrix and
$E(z,w)$ is the prime form.
Their explicit expressions
in term of the Schottky parameters
can be found in \cite{sc1}.
One can easily carry out the integration over $z$ in the above
expressions 
\be
{1\over 2\pi } \oint_0 dz \del X (z) f(z) = 2 \alpha'  p f(0) +\sqrt{2
\a'}  \sum_{n=1}
 \frac{a_n}{n!} \del^n f(z) |_{z=0}
\ee
and explicitly rewrite ${B^0}_I^\m$ and $C^0$ in terms of modes. In
particular, one finds here (and in the following \eq{B0}, \eq{B2}  and
\eq{higher}) terms containing only the zero modes $p$. In this case,
the arguments of the log's must be taken in absolute value.
This absolute value has its origin in the matrix element 
$D_{00}=(1/2) \log|\gamma'(0)|$ of the {\em real} infinite dimensional
representation of the projective group, relevant for the
{\em open} string (see ref. \cite{sc3}).

A couple of remarks are in order.
The modification in \eq{c} is the universal phase factor
depending on the external momentum. It corresponds to the Filk
phase \cite{filk} in the field theory limit.
The $\th$-dependent shift in \eq{b} corresponds in the field
theory limit to the shift
due to the dipole mechanism \cite{susskind0,zyin,susskind1}. The string
theory interpretation of this shift in terms of stretched string
was recently discussed in
\cite{hong}.
The shift in \eq{a} is new and, among other effects,
gives rise to  a $\th$ dependent measure factor for the zero modes of
the orbital degrees of freedom. One obvious advantage of our approach
is the clarity of the modifications due to $B$-field to the multiloop
string amplitudes: it is summarized in the zero mode contributions in
\eq{stringmaster}. As a result, apparently unrelated field theory
effects could find a uniform string explanation.

Carrying out the loop momentum integration, we obtain finally
\be \label{stringmaster}
V_{N;h}^\th = \widetilde{V}_{N;h}^{0}
\frac{1}{\sqrt{\det \frac{-A}{2} }} \exp
(-\frac{1}{2} B^T A^{-1} B +C) ,
\ee
where the determinant is taken over the space of Lorentz and loop
indices $(\mu I)$. When $\th=0$, it is
\be
\sqrt{\det (-A/2)} = (\det (-2 \pi i \a' \t))^{d/2}.
\ee
It is  remarkable that the multiloop string amplitudes
can again be written in terms of geometrical quantities of the Riemann
surface like the first Abelian differentials, period matrix and the
prime form, together with the intersection matrix of the loops.

Finally we close this section by writing down explicitly the
$\th$-dependences in the Reggeon vertex for
the  one loop case. First, there is  no modification to
${A^0}_{IJ}^{\m\n}$ and the modification to $C^0$ is
simply the usual field theory
Filk phase. Thus we concentrate on the effects from the shift
of ${B^0}^\m_I$. The one-loop  Abelian differential is
$ \omega = dz /z$.
With the convenient choice of $V_i(\r) = \r+ \r_i$,
we have
\be\label{B0}
 {B^0}_I^\m =
\sum_{r\in I_1 \cup I_2} \cB^{r\,\m} (\r) \ln(\r+\r_r),
\ee
where
\be\label{B1}
\cB^{r\,\m}(\r) = \frac{1}{2\pi} \oint_0 d\r \del X^{(r)\m}(\r)
\ee
is an operator in which the integration in carried out on any function
that multiply $\cB$ on the right.
Since $P=  \sum_{s\in I_2} p^s = \sum_{s\in I_2} \cB^s$ for the only
loop we have, it is
\be\label{B2}
B^{0\m} - i \th^{\m\n} P_\n = \sum_{r\in I_1 \cup I_2}
\cB^{r\,\m} (\r) \ln(\r+\r_r) -  \frac{i\th^{\m\n}}{2\a'}
\sum_{s\in I_2} \cB^s_\m .
\ee
Substituting into \eq{stringmaster}, we have
\bea \label{higher}
\frac{-1}{2A} B^2 &=& 2 \a' \cdot \frac{1}{8{\a'}^2 \ln k}
\Big( \sum_{\stackrel{r<s}{r,s\in I_1\cup I_2}}
\cB^r(\r) \cB^s(\r') \ln^2 (\frac{\r+\r_r}{\r'+\r_s}) \\
&+& \frac{i}{\a'} \sum_{\stackrel{r\in I_1}{s\in I_2}}
\cB^r(\r) \th \cB^s(\r') \ln [(\r+\r_r)(\r'+\r_s)] +
\frac{1}{4 \a'^2} \sum_{\stackrel{r\in I_1}{s\in I_2}}
\cB^r(\r) \th^2 \cB^s(\r')
\Big). \nn
\eea
The second line contains all the $\th$-dependence to the one-loop
string amplitude.
Putting
\be
\cB^r(\r) \to 2\a' p^r,
\ee
we recover the previous expression for the tachyon
amplitude and this is a consistency check of the correctness
of our results. However \eq{higher}  allows us to determine
the $\th$-dependence to the string amplitudes  for gluons as well
as any higher mass state. The
advantage of the Reggeon vertex formalism is obvious.

\sect{Noncommutative Field Theory Limit}

In this section, we consider the noncommutative $\Phi^3$ theory in
6-dimensions at two loops. We show that the 2-loop Reggeon vertex we
constructed in the previous section reproduces exactly the field theory
results. One-one correspondence between Feynman diagrams
and corners of string moduli space is established.

\subsection{Two Loops Noncommutative $\Phi^3$ in Six Dimensions}

\begin{figure}[ht]
\begin{center}
{\scalebox{1}{\includegraphics{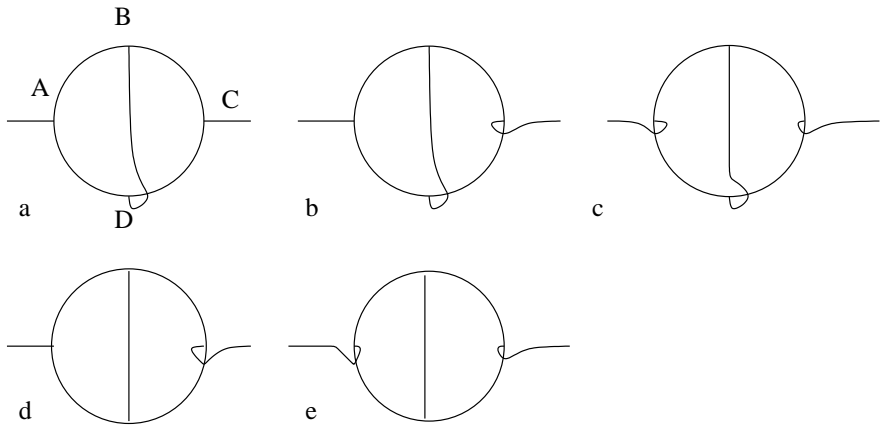}}}
\end{center} \caption{2-loop diagrams with external legs attached to
different propagators. }
\label{2loop1}
\end{figure}

Consider $\Phi^3$ interaction in $d=6$,
\beq\label{phi2l}
S_3 = \int \left({1\over 2}\, \partial_\m\phi\,\partial^\m\phi -
{1\over 2}\, m^2\phi^2 + {1\over 3!}\,
g_3~ \phi * \phi * \phi\right) d^dx ~.
\eeq
As an illustration,
we will consider 2-loop nonplanar diagrams with 2 external
legs.
The first type of diagrams have the two external legs attached to two
different propagators. Some of them are shown in Fig \ref{2loop1}.
\begin{figure}[ht]
\begin{center}
{\scalebox{1}{\includegraphics{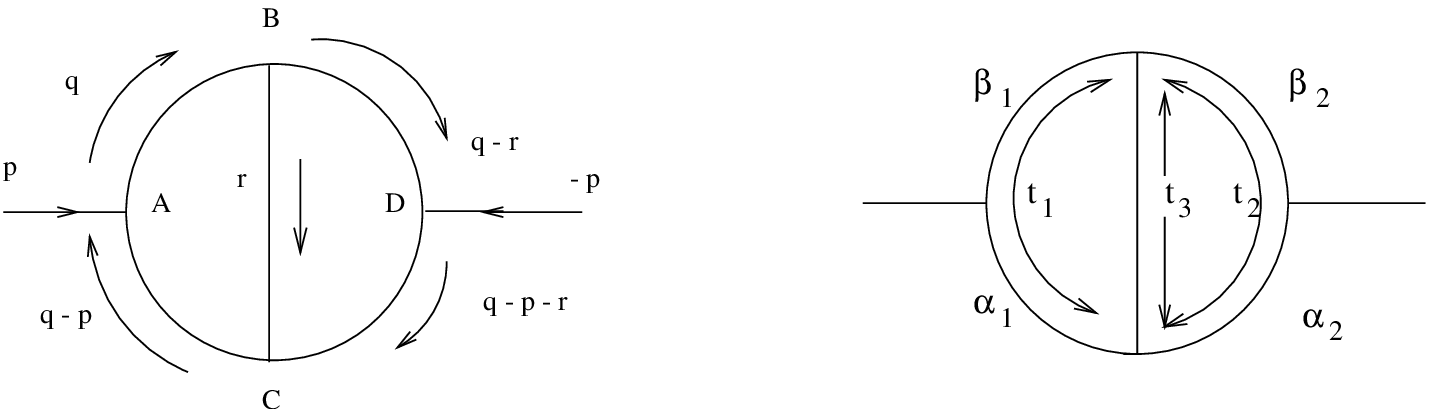}}}
\end{center} \caption{Momentum flow and Schwinger parameterization.}
\label{feycom}
\end{figure}

For all these diagrams, we have (up to overall numerical coefficient)
\be \label{field}
\cA = \int dq dr \int \cD \a \;\exp\left(
-\a_1 (q-p)^2 - \b_1 q^2 -\a_2(q-p-r)^2 -\b_2 (q-r)^2 -t_3 r^2
+ \varphi
\right),
\ee
where we have introduced Schwinger parameters and labelled the momenta
as in Fig \ref{feycom} with
\be
t_1 = \a_1+ \b_1, \quad t_2 = \a_2 +\b_2 .
\ee
$\varphi$ is the product of the phase factors arising at the various
junctions A, B, C, D due to the Moyal product and
\be
\int \cD \a = \int_{0}^{\infty} d t_3
\int_{0}^{\infty} dt_2  \int_0^{t_2} d\a_2
\int_{0}^{\infty} dt_1  \int_0^{t_1} d\a_1
e^{-(t_1+t_2+t_3)m^2}
\ee
is the integration over the Schwinger parameters. For convenience,
we have grouped the mass dependent factors here.
In this section, we will be interested in displaying the precise $\th$
dependence and thus will ignore overall numerical coefficients.

Let's start with Fig 4e. The phase factor is
\be
\varphi_{\it 4e} =- i p \th r.
\ee
Carrying out the $q$ integral, we obtain
\bea
\cA_{\it 4e} = \int dr \int \frac{\cD \a}
{(t_1+t_2)^{d/2}} \;
\exp \left( -p^2 [\a_1+\a_2 -\frac{(\a_1+\a_2)^2}{t_1+t_2}]
-\frac{1}{t_1+t_2}
[r^2 \Delta - 2 r^\m \xi_\m ]\right),
\eea
where
\be
\Delta := t_1 t_2 +t_2 t_3 +t_1 t_3
\ee
and
\be
\xi_\m := (\a_1 \b_2 -\a_2 \b_1)p_\m + \frac{i}{2}(t_1+t_2)(\th p)_\m.
\ee
Completing the square and integrate over momentum $r$, we obtain
\be
\cA_{\it 4e} = \int \cD \a \frac{1}{\Delta^{d/2}}
\exp \left(
p^2 \frac{ (\a_1 \b_2 -\a_2 \b_1)^2 }{(t_1+t_2)\Delta }
-p^2 [\a_1+\a_2 -\frac{(\a_1+\a_2)^2}{t_1+t_2} ]
- \frac{t_1 +t_2}{4\Delta} (\th p)^2
\right).
\ee
The exponent can be simplified easily and we obtain finally
\be \label{4e}
\cA_{\it 4e} = \int \cD \a \frac{1}{\Delta^{d/2}}
\exp \left( -p^2 [ \a_1+ \a_2 -
\frac{\a_2^2 (t_1+t_3) +  \a_1^2 (t_2+t_3) + 2\a_1\a_2 t_3}{\Delta} ]
- \frac{t_1 +t_2}{4\Delta} (\th p)^2
\right).
\ee
It is in this form that the field theory results may
be compared with string theory result most conveniently.

For the diagram of Fig 4b, we have the phase factor
\be
\varphi_{\it 4b} = i(p+r)\th q .
\ee
Carrying out the $r$ integral, we obtain
\be
\cA_{\it 4b} = \int dq \int \frac{\cD \a}
{l^{d/2}} \;
\exp \left( -p^2 [\a_1+\a_2 -\frac{\a_2^2}{l}]
-\frac{1}{l}[
q^\m \Dth_{\m\n} q^\n - 2 q^\m \xi_\m ]\right),
\ee
where
\be
l := t_2+t_3,
\ee
\be
\xi_\m := [\a_1 l + \a_2 t_3] p_\m + \frac{i}{2}(\a_2 - l) (\th p)_\m
\ee
and
\be
(\Dth)_{\m\n}:=\eta_{\m\n} \Delta - \frac{\th^2_{\m\n} }{4}~.
\ee
For these calculations it is convenient to choose $\th^{\m\n}$ to be a
block diagonal matrix; in this case $\Delta_\th$ is a diagonal matrix in
the Lorentz indices and can be easily inverted.
Completing the square and integrate over momentum $q$, we obtain
\be
\cA_{\it 4b} = \int \cD \a \frac{1}{\sqrt{\det \Dth}}
\exp \left(
{ \frac{1}{l} \xi_\m (\Dth^{-1})^{\m\n} \xi_\n
-p^2 [\a_1+\a_2 -\frac{\a_2^2}{l} ]}
\right) .
\ee
The exponent can again be simplified and we obtain finally
\bea \label{4b}
\cA_{\it 4b} = \int \cD \a \frac{1}{\sqrt{\det \Dth}}
\exp && \!\!\!\!\!\!\!\!\!\!
\left( - p [  \a_1+ \a_2 -
\frac{(t_1+t_3)\a_2^2 +  (t_2+t_3)\a_1^2 + 2\a_1\a_2 t_3}{\Dth}
]p \right) \nn\\
&& \times\exp [- (2 \a_2 -l) p \frac{\th^2}{4 \Dth}p ].
\eea

\begin{figure}[ht]
\begin{center}
{\scalebox{1}{\includegraphics{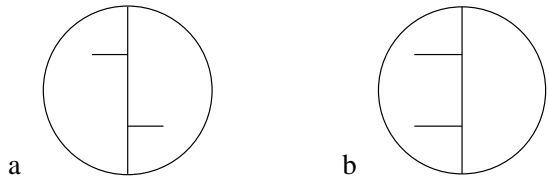}}}
\end{center} \caption{2-loop diagrams with external legs attached to
the same propagator. }
\label{2loop2}
\end{figure}

\begin{figure}[ht]
\begin{center}
{\scalebox{1}{\includegraphics{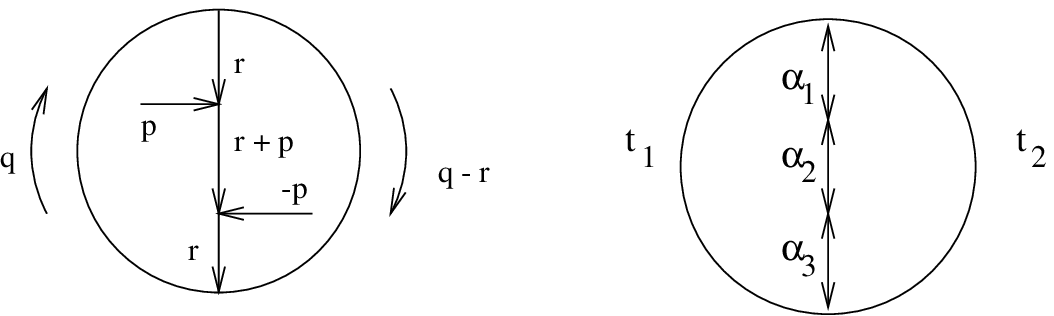}}}
\end{center} \caption{Momentum flow and Schwinger parameterization.}
\label{feycom2}
\end{figure}

Next we consider the type of diagrams (Fig \ref{2loop2})
with both external legs attached
to the same propagator. The momenta and Schwinger parameters are
labelled as in Fig \ref{feycom2}.

For these kind of diagrams, we have
\be
\cA = \int dq dr \int \cD \a \;\exp\left(
-t_1 q^2 -t_2 (q-r)^2 - \a_1 r^2 -\a_2(r+p)^2 -\a_3 r^2 + \varphi
\right),
\ee
where we have introduced
\be
\int \cD \a = \int_{0}^{\infty} dt_1 \int_{0}^{\infty}dt_2
\int_0^\infty dt_3 \int_0^{t_3} d\a_3 \int_0^{t_3-\a_3} d\a_1
e^{-(t_1+t_2+t_3)m^2},
\quad t_3:= \a_1 +\a_2 +\a_3
\ee
and $\varphi$ is a phase factor depending on the ``twisting'' at the
various vertices.
As an example, we  consider the diagram Fig \ref{2loop2} a. It is
\be
\varphi_{\it 6a} = i r \th p.
\ee
Evaluating the momenta integral, we obtain
\be \label{6a}
\cA_{\it 6a}  = \int \cD \a \frac{1}{\Delta^{d/2}} \exp \left(
-p^2 [\a_2 - \frac{\a_2^2}{\Delta}(t_1+t_2) ] -
\frac{t_1 +t_2}{4\Delta} (\th p)^2
\right).
\ee

Now we will turn to the string amplitude computations and their field
theory limits.

\subsection{Field Theory Limit of String Amplitudes}

Now we turn to the string formula. Two loops
amplitudes in commutative scalar theory has been considered in
\cite{russo2,russo2loop} where a precise one-one correspondence
(including numerical coefficients) between corners
of string moduli space and Feynman diagrams had been established.
Here we will be interested in reproducing
the $\th$-dependent terms of \eq{4e}, \eq{4b} and \eq{6a} from
\eq{stringmaster}.
The relation between the open string
coupling constant $g_{\rm op}$ and the field theory one $g_3$ has been
fixed at tree level~\cite{bcr}
\be\label{gthree}
g_3=2^{5/2} g_{\rm op} (2\a')^{d-6\over 4}~.
\ee
This equations allows us to write the string normalization in
the field theory quantities. Within field theory this normalization
depends on the different combinatoric factors of the various Feynman
diagrams and its determination can be quite cumbersome, in particular
in the non-commutative case.
Because of this, here we neglect all numerical factors also on the
string side and focus only on the integrand structure. However, at
least in the string derivation, numerical factors can be put back
easily.

In the general case of $h$ loops, the string
worldsheet is a $h$-annulus, which in the Schottky representation is
given by the $h$ generators $S_I$
\be
S_I (z) = \frac{a_I z+ b_I}{c_I z +d_I}, \quad a_I d_I - b_I c_I =1
\ee
of the Schottky group. Equivalently, one can introduce for each
generator the multiplier $k_I$ and the fixed points $\xi_I$ and
$\eta_I$, defined by
\be
\frac{S_I(z) -\eta_I}{S_I(z) -\xi_I} = k_I \frac{z-\eta_I}{z-\xi_I} .
\ee
As usual one can use the projective invariance to fix 3 of the $2h$
fixed points and inequivalent
$h$-loop worldsheet is parametrized by $2h-3$ fixed points and $h$
multipliers. To arrive at a noncommutative field theory,
in addition to the double scaling limit \eq{double}, one also has to
go to corners of the string moduli space in order to get a finite
contribution to the string amplitude that can be identified with
Feynman diagrams.

\begin{figure}[ht]
\begin{center}
{\scalebox{1}{\includegraphics{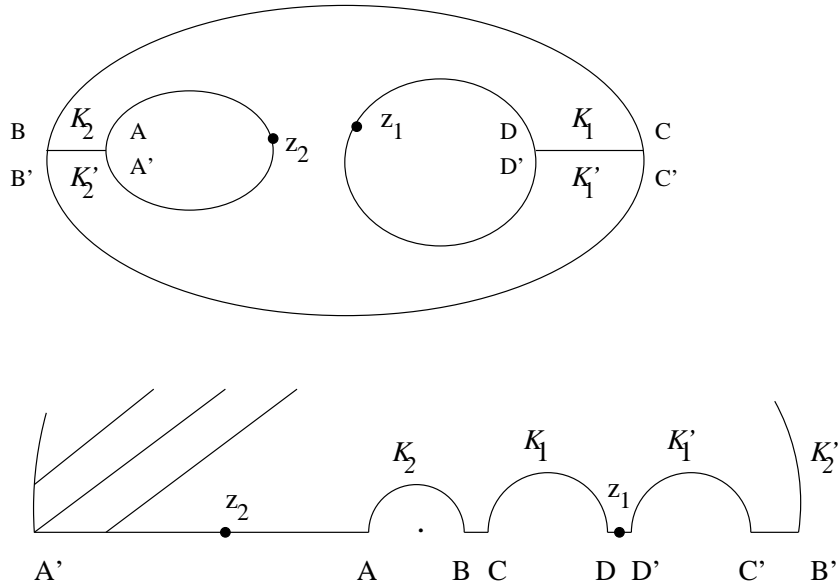}}}
\end{center} \caption{Two-annulus  and the Schottky representation}
\label{scho2loop}
\end{figure}

For the case of two loops, the string worldsheet is a two-annulus
which in the Schottky representation is represented by
the shaded region in Fig \ref{scho2loop}. We
can fix $\eta_2=0, \xi_1=1$ and $\xi_2=\infty$ using the projective
invariance and we are left with one fixed point $\eta_1$ and two
multipliers $k_1, k_2$. One can verify that \cite{russo2}
\bea
B = -A = \sqrt{k_2} , \quad
C= \frac{\eta_1- \sqrt{k_1}}{1- \sqrt{k_1}}, \quad
D=  \frac{\eta_1+ \sqrt{k_1}}{1+\sqrt{k_1}}, \nn\\
B' =-A' = \frac{1}{\sqrt{k_2}}, \quad
C' = \frac{1- \eta_1 \sqrt{k_1}}{1- \sqrt{k_1}}, \quad
D' =  \frac{1+ \eta_1 \sqrt{k_1}}{1+ \sqrt{k_1}}.
\eea
The irreducible vacuum bubble represents the backbone for all field
theory diagram we are interested in. It is obtained, from the string
formula, in the limit $\a' \to 0$, $k_1, k_2, \eta_1 \to 0$ with the
following $t_i$'s fixed
\be\label{fp-st}
-\a' \ln k_1 = t_1 +t_3 , \quad -\a' \ln k_2 = t_2+t_3, \quad
-\a' \ln \eta_1 = t_3~,
\ee
where, as usual, the $t_i$'s are identified with the field
theory Schwinger parameters introduced in the previous section.
In this limit, the matrix $A^0$ and the 2-loop period matrix is
given by
\be \label{basicA}
-\frac{1}{2}A^0 = -\a' \; 2\pi i \tau {\bf 1} =
\left(
\begin{array}{cc}
(t_1 + t_3){\bf 1}  & t_3{\bf 1}   \cr
t_3{\bf 1}  & (t_2+ t_3) {\bf 1}
\end{array}
\right) ,
\ee
where we have written $A^0$ as a $2\times 2$ block (will be $h\times h$
block for $h$ loops case) with
$ {\bf 1}$  being the $d\times d$ identity matrix in the Minkowski space.

In the string world-sheet the external states are represented by
punctures on the borders of the surface: in our case, we introduce
the first external state in $z_1$ and the second one in $z_2$.
In the field theory limit, these points are identified with remaining
two Schwinger parameters $\a_i$ through a relation similar to that of
\eq{fp-st}. However, the exact form of the identification between
puncture and Schwinger parameters depends on the particular corner of
the $z$'s integration region one is looking at; technically, this is
the reason why a single string Green function can give rise to field
theory diagrams having different forms in terms of Schwinger
parameters.
Following \cite{russo2,russo2loop}, we start looking at the region
where $z_2\to A'$ and $z_1\to D'$ which is related to the field theory
diagram of Fig. 4e. In this case, $B^0$ is given in terms of the
Schwinger parameters $\a_i$'s by
\be
\frac{1}{2}\left(
\begin{array}{c}
{B^0}^\m_1 \cr
{B^0}^\m_2
\end{array}
\right)
= \left(
\begin{array}{c}
\a_1 p^\m \cr
- \a_2 p^\m
\end{array}
\right) .
\ee

\begin{figure}[ht]
\begin{center}
{\scalebox{1}{\includegraphics{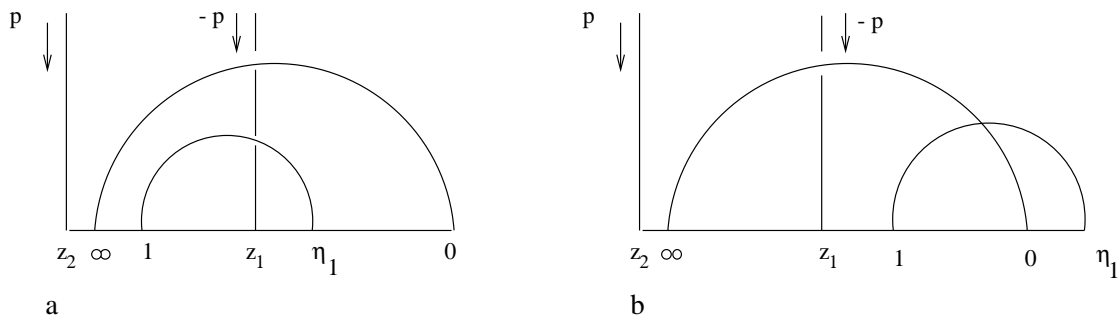}}}
\end{center} \caption{String worldsheets corresponds to field theory
diagrams Fig. 4e, 6a and 4b. }
\label{worldsheet1}
\end{figure}

Now we are ready to reproduce  the field theory results from
\eq{stringmaster}.
The  string worldsheet
that gives a nonzero contribution  to the field theory diagram
Fig 4e is given by  Fig \ref{worldsheet1}a.
Since there is no intersection between the two loops, the only
modification is in $B_I$. We have
\be
{(-A^0/2)^{-1}} =
\frac{1}{\Delta}
\left(
\begin{array}{cc}
(t_1 + t_3){\bf 1}  & -t_3 {\bf 1} \cr
-t_3{\bf 1}  & (t_2+ t_3){\bf 1}
\end{array}
\right)  .
\ee
and
\be\label{Bconf2}
B_I^\mu ={B^0}^\m_I + \frac{i}{2} (\th p)^\m \left(
\begin{array}{c}
1 \cr 1
\end{array}
\right) .
\ee
Thus, after the integration over the momenta running in the loops, the
string formula gives the following $\th$ dependent part
\be
-\frac{1}{2} B^T A^{-1} B  =
\frac{(t_1+t_3)\a_2^2 +  (t_2+t_3)\a_1^2 + 2\a_1\a_2 t_3}{\Delta}p^2
-\frac{1}{4\Delta} (\th p)^2 (t_1+t_2).
\ee
Together with measure factor  $\Delta^{-d/2} $ and
the other pieces coming from $ \tilde{V}_{N;h}^{0}$ and $C_0$,
the $\th$ independent piece of \eq{stringmaster}
is exactly the same as the field theory
integrand \eq{4e}. Moreover, the $\th$-dependence is also precisely
reproduced.

A remark about the integration region is in order. As explained in
\cite{russo2loop}, the corner of $z_i$'s considered yields the correct
functional form of Feynman diagram in Fig. 4e, but does not reproduce
the field theory integration region over the Schwinger
parameters. However, we know that a field theory diagram is identified
with the sum of contributions from a few different corners of the
string moduli space. This sum reconstructs the expected
field theory integration region,
without modifying the integrand previously found.
Since turning on a $B$-field does not change this identification,
the situation is therefore the  same as in \cite{russo2loop}. It is easy
to check that the complete field theory result \eq{4e} is reproduced
from string theory.

Next we consider the string worldsheet configuration
of Fig \ref{worldsheet1} b. In this case,
\be
-A/2 =
\left(
\begin{array}{cc}
(t_1 + t_3) {\bf 1} & t_3 {\bf 1} + \frac{i}{2} \th \cr
 t_3 {\bf 1} - \frac{i}{2} \th & (t_2 + t_3) {\bf 1}
\end{array}
\right) ,
\ee
where again we have written $A$ as a $2\times 2$ block with
$ {\bf 1}$ and $\th$ being $d\times d$ matrices.
It is easy to work out the inverse,
\be
(-A/2)^{-1} =
\frac{1}{\Delta_\th}
\left(
\begin{array}{cc}
(t_2 + t_3) {\bf 1} & -t_3 {\bf 1} - \frac{i}{2} \th \cr
- t_3 {\bf 1} + \frac{i}{2} \th & (t_1 + t_3) {\bf 1}
\end{array}
\right) .
\ee
As for $B$, it is
\be
\frac{1}{2}\left(
\begin{array}{c}
B_1^\m \cr
B_2^\m
\end{array}
\right)
= \left(
\begin{array}{c}
\a_1 p^\m + \frac{i}{2} (\th p)^\m \cr
- \a_2 p^\m
\end{array}
\right).
\ee
Therefore
\be
-\frac{1}{2} B^T A^{-1} B =
p \frac{(t_1+t_3)\a_2^2 +  (t_2+t_3)\a_1^2 + 2\a_1\a_2 t_3}{\Dth}p
+ (t_2+t_3- 2 \a_2 ) p \frac{\th^2}{4 \Dth}p.
\ee
Again the complete field theory result \eq{4b} including the region of
integration is reproduced.

Finally we consider the field theory diagram Fig. 6a.
As explained in \cite{russo2}, Fig. \ref{worldsheet1}a actually
contributes to both types of field theory diagrams: Fig. 4e
with external legs in distinct propagators and Fig. 6a with external
legs on the same propagator. One has to be careful in selecting the
range of the punctures in order to get the desired correspondence. For
the present case of Fig.~6a, one has to consider the range
$z_2\in[-1,-\eta]$ and $(z_1-\eta)/(z_1-1)\in[-1,-\eta]$.
According to \eq{a}-\eq{c},
$A$ is unmodified from $A_0$ and is given by \eq{basicA} since
there is no intersection of the loops.
As for $B$, it is given by \eq{Bconf2}.
The only difference from the case of Fig. 4e is in the range of the
punctures $z_1,z_2$ and hence a different expression of $B^0$
\be
\frac{1}{2}\left(
\begin{array}{c}
{B^0}_1^\m \cr
{B^0}_2^\m
\end{array}
\right)
= \left(
\begin{array}{c}
\a_2 p^\m \cr
\a_2 p^\m
\end{array}
\right)~.
\ee
For our purpose, we concentrate on the $\th$
independent terms.
Precisely the $\th^2$ term in \eq{6a} is reproduced and confirm
the complete field theory result.

We have also checked the other field theory
diagrams in Fig 4 and 6
and they are all reproduced correctly from  the string
master  formula \eq{stringmaster}.

\sect{Conclusions and Discussions}

In this paper, we used the open string operator formalism to construct
multiloop string amplitudes with $B$-field.
The key observation is that when expressed in terms of the commutation
relations of the string modes, the effects of noncommutativity is very
simple: apart from an overall scaling of the operators, only
the zero mode commutation relation is modified.  This allowed us to
repeat the basic steps in operator formalism and construct multiloop
string amplitudes with $B$-fields.
In particular, we constructed the open string vertex operators
and determined directly the one-loop open string Green
function. For higher loop amplitudes,
we constructed the multiloop Reggeon vertex
by  sewing together legs of a tree level Reggeon vertex.
It was thus possible to write down explicitly
string amplitudes with $B$-field with any number of loops.
Moreover one-one correspondence between corners of open
string moduli space
and Feynman diagrams is again established.

With the master formula at our disposal, it is in principle possible to
tackle many field theory questions,
which may appear mysterious otherwise,
from the string point of view.
This is one major advantage of the string approach.
One such question is the
UV/IR effect in  noncommutative field theory.
Since the $\th$ dependences in any higher loop amplitude are isolated
explicitly in our master formula \eq{stringmaster}, by taking the
field theory limit of the string result, one can
do a systematic analysis of  the UV/IR effects
in a noncommutative field theory at arbitrary loop levels.
Thus this approach may provide an uniform
string theory understanding of the UV/IR effect.

Another possible application is in the understanding of a similar
effect in 3-dimension. It was found in \cite{cs} that a  Majorana fermion
induces a noncommutative Chern-Simons action at one loop. The
coefficient is a step function which is $1/2$ when the theory is
noncommutative  and  is 0 when the theory is commutative.
It would be interesting to understand this as well as other
perturbative aspects of this theory from the string
point of view.

Although we did not write down explicitly the open string Green function
for higher loop, it is easy to extract it from the master formula
\eq{stringmaster}.
It is also easy to generalize the considerations of this paper to the
superstring case and to put Chan-Paton factors to obtain results for
noncommutative field theory with $U(N)$ gauge group.

So far, noncommutative D-branes appear in string theory in two basic
forms: Moyal-deformation in the case of flat spacetime with
$B$-field, and ``fuzzy deformation'' in the case of D-branes in WZW
model. In particular the latter is generically a more general
nonassociative deformation \cite{ARS1}.
It would be interesting to explore other possibility of
noncommutative geometry and deformation that can arise in string
and M-theory.

We hope to return to some of these issues in the future.

\vspace{.5cm}

\noindent{\large \bf Acknowledgments}

This work was partially supported by the Swiss National Science
Foundation, by the European Union under TMR contract
ERBFMRX-CT96-0045, by the Swiss Office for Education and
Science and by MURST (Italy).

\ed